\documentclass[
  journal=pasa,
  manuscript=research-paper, 
  year=2020,
  volume=37,
]{cup-journal}

\usepackage{microtype,siunitx,booktabs, rotating}

\usepackage[frozencache,cachedir=.]{minted} 

\title{Efficient Summation of Arbitrary Masks $-$ ESAM}

\author{V. Gupta}
\affiliation{Australia Telescope National Facility, CSIRO, Space and Astronomy, PO Box 76, Epping, NSW 1710, Australia}
\email[V. Gupta]{vivg269@gmail.com}

\author{K. W. Bannister}
\affiliation{Australia Telescope National Facility, CSIRO, Space and Astronomy, PO Box 76, Epping, NSW 1710, Australia}

\author{C. Flynn}
\affiliation{Center for Astrophysics and Supercomputing, Swinburne University of Technology, Post Office Box 218, Hawthorn, VIC 3122, Australia}
\alsoaffiliation{ARC Centre of Excellence for Gravitational Wave Discovery (OzGrav), Post Office Box 218, Hawthorn, VIC 3122, Australia}

\author{C. James}
\affiliation{International Centre for Radio Astronomy Research, Curtin University, Bentley, WA 6102, Australia}

\received {dd Mmm YYYY}
\revised  {dd Mmm YYYY}
\accepted {dd Mmm YYYY}
\published{01 January 2023}


\begin{document}

\begin{abstract}

Searches for impulsive, astrophysical transients are often highly computationally demanding. A notable example is the dedispersion process required for performing blind searches for Fast Radio Bursts (FRBs) in radio telescope data. We introduce a novel approach --- Efficient Summation of Arbitrary Masks (ESAM) --- which efficiently computes 1-D convolution of many arbitrary 2-D masks, and can be used to carry out dedispersion over thousands of dispersion trials efficiently. Our method matches the accuracy of the traditional brute force technique in recovering the desired Signal-to-Noise ratio (S/N) while reducing computational cost by around a factor of 10. We compare its performance with existing dedispersion algorithms, such as the Fast Dispersion Measure Transform (FDMT) algorithm, and demonstrate how ESAM provides freedom to choose arbitrary masks and further optimise computational cost versus accuracy. We explore the potential applications of ESAM beyond FRB searches.

\end{abstract}

\section{Introduction }
\label{sec:int}


Many astrophysical sources emit short-duration radio pulses, such as pulsars \citep{Hewish1968}, fast radio bursts \citep{Lorimer2007}, rotating radio transients (RRATS) \citep{McLaughlin2006_RRAT} and ultra-long period sources \citep{Caleb_ULP2022_Nature, HurleyWalker2023Nature_ULP}. Sensitive searches for new transient sources are highly motivated by a number of astrophysical questions, such as measuring the cosmological constant or cosmic baryon density with fast radio bursts, to probing gravity and the neutron star equation of state with pulsars.

As a radio pulse propagates through cold ionised plasma, it undergoes a frequency-dependent time delay, known as ``dispersion'', such that

\begin{equation}
    \label{Eq:dispersion eqn}
    \Delta t = K\,{\rm DM} \,\left(f_{1}^{-2} - f_{2}^{-2}\right),
\end{equation}

where $\Delta t$ is the time delay, $f_1$ and $f_2$ are lower and upper frequencies, and $K$ is a combination of physical constants ($K=\frac{8 \pi^2 \epsilon_0m_ec}{e^2} {\rm pc}^{-1}$, \cite{Kulkarni2020_DM}). DM is the ``dispersion measure'' --- or the integrated electron density $(n_e)$ along the line-of-sight from the source at distance $D$ to the receiver:

\begin{equation}
    {\rm DM} = \int_{0}^{D} n_e(l) dl.
\end{equation}  

Typical radio telescope processing will sample the electric field from a telescope, form $n_c$ equally-spaced frequency channels, and integrate the power over an integration time $T$. This produces a sequence of sky power as a function of time and frequency, known as a dynamic spectrum. While FRBs must be emitted via coherent radiation processes \citep{Petroff2019_review}, their emission mechanisms are not understood well enough to be predictive about the signal shape in the voltage domain, hence matched filtering in the power domain is commonly used for searching for FRBs.

Optimal detection of a signal embedded in noise in the dynamic spectrum requires a matched filter. In the case of a dynamic spectrum, a matched filter requires time-convolution (multiply and sum for multiple time samples) of the input signal with a 2-D mask that matches the expected signal shape. When searching blindly for new transient sources, the DM of the signal shape cannot be known {\it a priori} and surveyors need to convolve their data with a bank of masks with different DMs - a process known as dedispersion. The dedispersion masks are based on the dispersion Equation (\ref{Eq:dispersion eqn}). 
Carrying out dedispersion over thousands of dispersion trials is a highly computationally expensive operation, and usually, a trade-off between accuracy and computational speed has to be made.
In this paper, we describe an improved algorithm for dedispersion both in speed and recovery of the astrophysical signal, allowing computational resources to be allocated to a wider search parameter space or other tasks.  

\subsection{Sources of S/N loss}

The maximum theoretical signal-to-noise (also known as the matched-filter S/N) of a signal $x_i$ convolved with a mask with weights $w_i$ is obtained using assuming independent Gaussian noise with unit variance, is given by:

\begin{equation}
({\rm S/N})_{\rm matched filter} = \frac{\sum_i x_i w_i}{\sqrt{\sum_i w_i^2}}.
\end{equation}

Here we have assumed S/N as a ratio of mean to the standard deviation of the measured signal, rather than a ratio of powers.

All practical dedispersion implementations perform either an implicit or an explicit 1-D convolution of the dynamic spectrum with a bank of masks along the time axis. Practical algorithms do not often achieve the matched filter S/N due to a mismatch between the measured pulse and the assumed masks. There are several reasons for this mismatch. 

Firstly, resource limitations always cap the number of dispersion trials which can be searched, or equivalently, the number of masks that can be convolved. Typically $\sim 10^3$ dispersion trials spanning a wide DM range are used, with gaps in DM between trials. A pulse which falls in the gap between adjacent trials will have an S/N lower than a perfectly matched pulse, an effect known as ``scalloping'' \citep{Keane-Petroff2015}. 

Secondly, many algorithms simply add up the set of cells with equal weights and do not perform an explicit 1-D convolution with arbitrary (continuous-valued) kernels. This is equivalent to 1-D convolution with a binary mask. This results in loss of S/N because the input is continuous-valued but the kernel is binary-valued. 

Thirdly, the structure of some algorithms (particularly tree-based algorithms) may not approximate the $f^{-2}$ dependence on the delay very well. We describe this shape mismatch as the ``accuracy'' of the mask. This results in loss of S/N again because the effective mask does not match the signal. 

Finally, measured pulses exhibit a wide range of morphologies in addition to dispersion. They may be caused by astrophysical effects (such as scattering and/or scintillation) and instrumental effects (such as dispersion smearing due to limited frequency resolution). Typically, accounting for all of these effects would require a prohibitively large number of kernels, so the designer must compromise and choose a set of kernels that samples the parameter space as efficiently as possible, within the computing budget. This results in a fundamental trade-off between S/N and computational resources.

\subsection{Practical dedispersion algorithms}

Many dedispersion algorithms have been developed over the years. The algorithms fall broadly into 2 classes. Firstly, there are ``brute force'' methods that effectively shift and add channels to follow a set of dispersion tracks as defined by Equation~\ref{Eq:dispersion eqn}. Such algorithms have a complexity of $\mathcal{O}(N_c N_d)$, where $N_c$is the number of channels, and $N_d$ is the number of dispersion trials. Secondly, ``fast'' or tree-based algorithms exploit the fact that adjacent dispersion trials share many common partial-sums, and re-use these partial-sums to reduce the computational cost. Fast algorithms have a complexity of $\mathcal{O}(N_d log_2{N_c})$. 

Choosing a dedispersion algorithm typically offers a trade-off between computational cost, which favours tree-based algorithms, and accuracy (which translates to S/N and ultimately detection rate) which favours brute force algorithms.

With the advent of high-performance parallel and accelerated computing, brute force dedispersion algorithms have been adapted to work on parallel computing systems such as on GPUs in HEIMDALL \citep{Barsdell_HEIMDALL} and in AMBER \citep{Sclocco2016_IEEE_AMBER_dedispersion_tuning}, achieving very substantial speed-up in dedispersion, and thus wider parameter space searches within a given performance budget. 

\cite{Taylor1974} introduced the tree dedispersion algorithm, which uses a more computationally efficient technique of divide-and-conquer \citep{Rajwade2024_frb_search_algo_review}. This algorithm has the limitation that it only supports linear dispersion tracks. Quadratic dispersion tracks can be achieved by carefully padding the input channels with zeros before transforming. 

More recently, \cite{Zackay2017_FDMT} developed a significant new tree-based algorithm, called the ``Fast Dispersion Measure Transform'' (FDMT). The FDMT approximates quadratic dispersion tracks in time-frequency data, without the need for additional padding. FDMT is being used in a number of on-going transient detection pipelines to search for transient signals efficiently \citep{Bannister2017, Mandlik2024_UTMOST_NS}.

In this paper, we introduce a new tree-based algorithm --- ESAM, or ``Efficient Summation of Arbitrary Masks''.  
In Section \ref{sec:ESAM}, we describe the structure and the working of the algorithm. In Section \ref{sec:ESAM_perf}, we describe the methods used to quantify its performance, and present the results in Section \ref{sec:results} placing it in the context of the performance of brute-force and FDMT dedispersion algorithms. We present our concluding remarks in Section \ref{sec:discussion}.

\section{Efficient Summation of Arbitrary Masks - ESAM}
\label{sec:ESAM}

We call our method the Efficient Summation of Arbitrary Masks (ESAM). ESAM is a generic algorithm for efficiently computing 1-D convolutions of a bank of arbitrary 2-D masks. For the purposes of dedispersion, the convolution direction is the time axis, the sum direction is over the channel axis, and the bank of masks comprises shapes of dispersed pulses at the desired DM trials.  

The structure of ESAM is inspired by the tree structure of the FDMT. The key difference is that the sums performed by ESAM are driven by lookup tables, whereas the FDMT computation is fixed by an equation. The ESAM lookup tables are prepared very intuitively. The designer generates a bank of arbitrary 2-D masks. The masks can account for any astrophysical or instrumental effects they choose (e.g. dispersion, scattering, intra-channel smearing). The designer can choose arbitrary spacing of DM trials between consecutive templates. The designer then loads the 2-D masks sequentially into the ESAM tree. As each mask is loaded, the ESAM tree stores the 2-D mask with an encoding that minimises the number of redundant operations with respect to previously loaded masks. When evaluating against incoming data, a full 1-D time-convolution is performed with each mask across time. The sums across frequency are performed with no redundant operations. 

ESAM computes every user-supplied mask with 100\% accuracy, i.e. its S/N performance is identical to any brute force algorithm for which the 2-D masks can be specified numerically. ESAM guarantees computation of the convolution with no redundancy i.e. it guarantees maximum re-use of partial sums between different kernel outputs and no redundant convolutions are computed. Therefore, ESAM will always use fewer operations than a brute-force algorithm. ESAM gives the designer complete freedom to choose arbitrary convolution kernels and trade computational cost versus accuracy. 

ESAM has no internal concept of dispersion or pulse shape as such --- it simply performs convolution over a specified map and keeps track of different output ``products''. We therefore use the ``product'' terminology in the description below. Products may map to dispersion trials, pulse types (e.g. Gaussian, exponential), pulse widths, or spectral patterns, at the designer's choice.

\subsubsection{The pseudocode and reference Python implementation}

Our functional Python implementation of ESAM is available at \url{https://github.com/vivgastro/ESAM}. This implementation makes use of classes to aid with readability and comprehension. This implementation should be considered the primary reference describing how the algorithm works. It includes sanity checks and other functionality.
To aid the reader, we have written key algorithms in pseudocode with a Python-like syntax. For the sake of brevity and clarity, these listings do not match the Python reference implementation exactly, omitting some details.

\subsubsection{The trace representation of a 2-D mask}

\begin{figure*}
    \centering
    \includegraphics[width=0.75\linewidth]{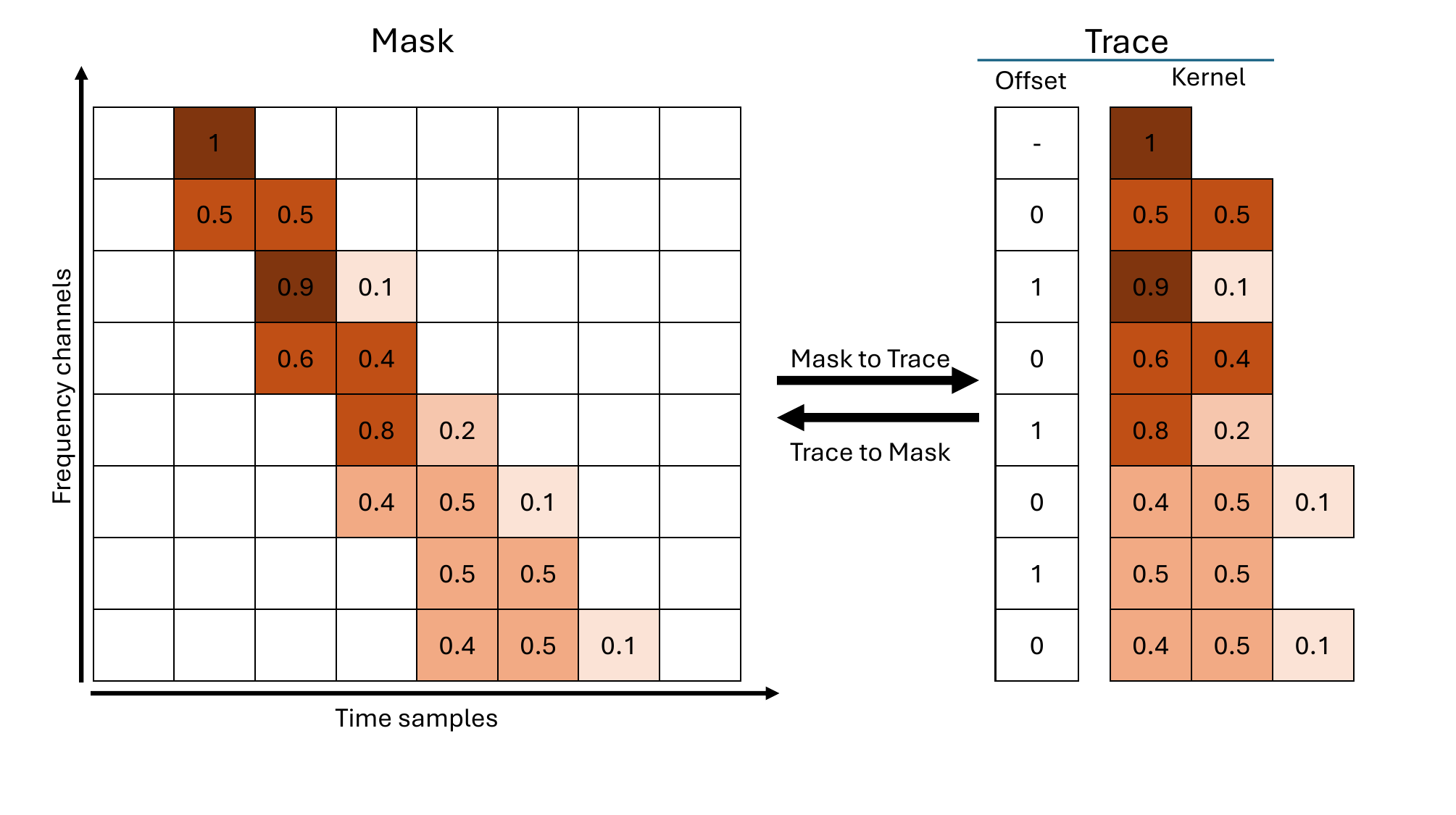}
   \caption{ESAM represents an arbitrary 2-D mask as a ``trace''. The trace is comprised of 2 structures: a set of relative offsets between a channel and the preceding, and the 1-D convolution kernels themselves without the leading or trailing zeros. All masks must be converted to traces before they can be given to the ESAM tree.}
    \label{fig:trace}
\end{figure*}

Before describing ESAM, we need to describe a data structure which we call a ``trace''. A trace is an alternative representation of a 2-D mask, or dispersion trial. The trace representation facilitates computations in the ESAM code when generating lookup tables for a given mask. The mapping between a 2-D mask and a trace, and the method for splitting a trace, is shown in Figure~\ref{fig:trace}.

The trace representation takes the 2-D mask and produces two data structures. The ``offsets'' array is the number of initial zero samples between the channel of interest, and the preceding channel. The offset of the top channel is undefined. The second structure is a list of the 1-D kernel weights for each channel, with the leading and trailing zeros removed.
In Figure~\ref{fig:trace} we show the procedure for making a trace from a mask.

\subsection{Tree structure of ESAM}
\label{subsec: ESAM tree structure}

\begin{sidewaysfigure*}
    \centering
    \includegraphics[width=0.95\linewidth]{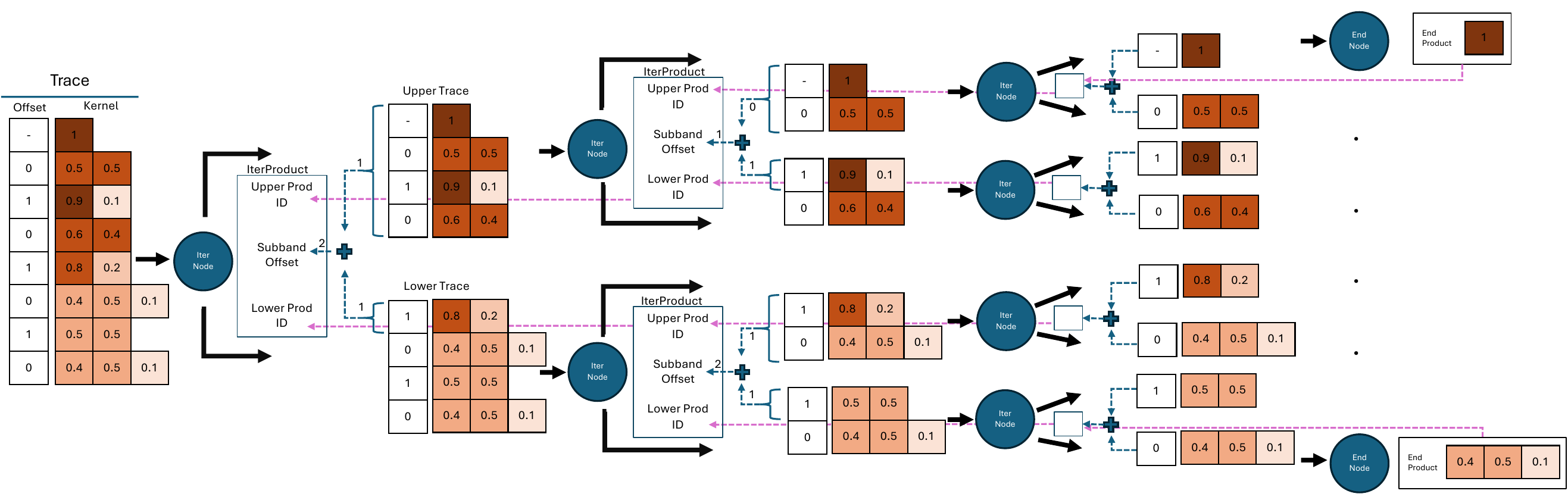}
   \caption{Diagram showing how a trace is digested while building an 8-channel ESAM tree. The trace is split into upper and lower halves. The subband offset is computed and saved and the trace halves are sent to the respective child nodes. This proceeds until the 1-D kernels are saved by the EndNodes as EndProducts. Each node returns the index (product ID) in its lookup table to its caller. The internal (IterNodes) save the upper and lower product ID, and the subband offset in a lookup table called the IterProduct. This lookup table is used in the evaluation stage. A procedural description of the building of a given IterNode and EndNode is depicted in Figures~\ref{fig:Iternode precodure} and \ref{fig:Endnode procedure}, respectively. Please see Section~\ref{subsec: ESAM tree procedure} for a more detailed discussion.}
    \label{fig:tree_building}
\end{sidewaysfigure*}

To aid with the explanation of ESAM, and with understanding our Python implementation, we first describe the algorithm in terms of a full binary tree as shown in Figure~\ref{fig:tree_building} For $N_c$ which is not a power of 2, the binary tree will not be full, and extra care constructing the tree is required, but the algorithm remains essentially unchanged. 

The ESAM tree is built by feeding in the bank of traces one by one, starting from an empty tree. We describe below the structure of an ESAM tree (the tree-building procedure is found in subsection~\ref{subsec: ESAM tree procedure}).
The top layer (or root node) processes all channels provided in the trace. Each lower layer contains twice as many nodes as the layer above. Nodes in a given layer processes 2$\times$ fewer channels than a node in the layer above. The root node and internal nodes are of type ``IterNode''. IterNodes in the same layer process disjoint sub-bands. The leaf nodes, which are of type ``EndNodes'' processes a single channel. An ESAM tree processing $N_c$ channels (power of 2), will have $N_c$ leaf nodes and $log_2{N_c}$ layers. Layer $i$ contains $2^i$ nodes, with each node processing data for a total of $2^{log_2{N_c}-i}$ channels.

Each node maintains several attributes. IterNodes maintain references to two child nodes, and a lookup table defining how to compute output products given input products from the child nodes. The lookup table is a list of 3-tuples, which we call IterProducts: `(upper\_product\_id, lower\_product\_id, subband\_offset)'. The `subband\_offset' represents the total offset between the upper and the lower subbands given to the current IterNode, and is computed as the sum of all offsets in the upper half of the trace, and the offset of the top channel in the lower half of the trace. Each EndNode maintains a list of 1-D convolution kernels as EndProducts. An EndNode applies its convolution kernels to the channel for which it is responsible. 

At design time, the designer loads traces into the tree. The preparation procedure uses the traces to compute the lookup tables. We call this `preparing' the tree. At run time, we present a time-frequency block to the tree, which computes the bank of convolutions. We call this `evaluating' the tree.

The general structure, as well as the process for preparing the tree from a single trace, are shown in Figure~\ref{fig:tree_building}.

\subsubsection{ESAM preparation procedure}[htb!]
\label{subsec: ESAM tree procedure}
\begin{figure*}
    \centering
    \includegraphics[width=0.95\linewidth]{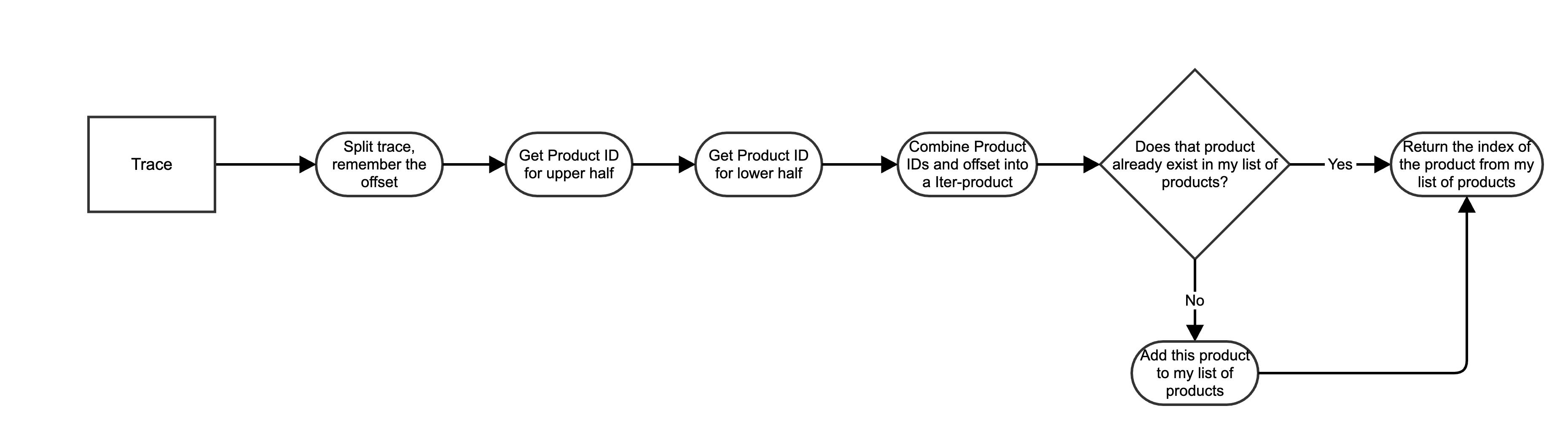}
    \caption{ESAM IterNode building procedure.}
    \label{fig:Iternode precodure}
\end{figure*}

In the preparation stage, a trace is fed into the root node, which is split up recursively from top to bottom into children nodes. All nodes update their lookup tables if necessary and return the index in their lookup table, known as a ``product ID'' that identifies the trace. Once a trace is loaded, loading the same trace again will return the same product ID. The procedure is illustrated in Figure~\ref{fig:tree_building} and proceeds as follows. The designer prepares a 2-D mask and converts it to a trace. The trace is given to the root node. The root node splits the trace into upper and lower halves, and saves the subband offset between the two halves as shown in Figure~\ref{fig:tree_building}. It passes the upper sub-trace to the upper child node and the lower sub-trace to the lower child node. This process is repeated recursively by each IterNode in a top-down fashion. The recursion terminates when a single channel trace, i.e. a 1-D kernel is presented to an EndNode.

The recursion now unwinds upwards. Each EndNode checks whether the supplied 1-D kernel is already in its list of kernels (i.e. its list of EndProducts). Two 1-D kernels are deemed the same if they have identical coefficients. If the kernel has not yet been saved, it appends the 1-D kernel into its list of EndProducts. The EndNode returns the index, or product ID, of the supplied 1-D kernel to its parent. Every 1-D kernel in an EndNode lookup table is unique. 

Each parent IterNode receives the two ProductIDs from its respective child nodes, combines them with the subband offset saved in the downward recursion, and forms the 3-tuple IterProduct. Two IterProducts are deemed to be the same if all three attributes of the 3-tuple IterProduct (`upper\_product\_id', `lower\_product\_id', `subband\_offset') are equal. Once again, if this IterProduct is not present in its lookup table, it is appended to the lookup table (or its list of IterProducts). Every IterProduct in the IterNode lookup table is unique. The IterNode returns the index in the lookup table of the IterProduct. This process unwinds upwards until the root node returns the ProductID of the IterProduct in its own lookup table. The ProductID returned by the root node is the index in the output data where the result of the 1-D convolution of the input data with the trace that was just supplied, is found.

\begin{figure*}[htb!]
    \centering
    \includegraphics[width=0.75\linewidth]{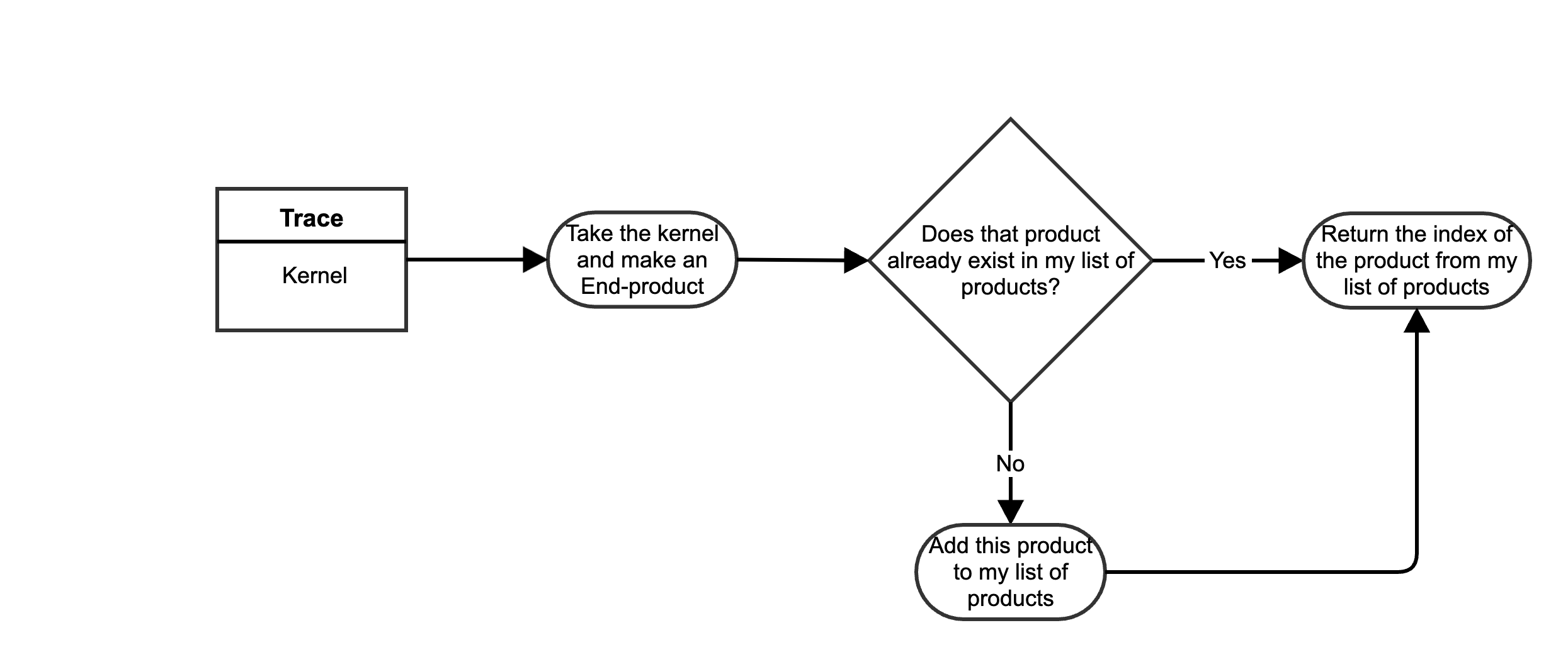}
    \caption{ESAM EndNode building procedure.}
    \label{fig:Endnode procedure}
\end{figure*}

A key feature of ESAM algorithm is that before each node adds a new product to its lookup table, it checks if this product matches with any of the existing products already present in the lookup table. If found, the node returns the index of the existing product. If not, it adds the new product to the list and returns the index of the newly added element. In computing, this is known as memoization. Memoization of the mapping between a trace and a product ID in all nodes in the tree is the way ESAM guarantees that no redundant partial sums, and no redundant convolutions are performed.

Pseudocode for IterNode and EndNode function dispatching is shown in Listing~\ref{listing:types}. The IterNode preparation pseudocode shown in Listing~\ref{listing:prep_iternode}, and the EndNode preparation pseudocode is in Listing~\ref{listing:prep_endnode}.

\subsubsection{ESAM Evaluation procedure}

Once the ESAM tree has been built with the set of traces, it can be used at run time to process blocks of dynamic spectra data of shape ($N_c, N_t$). Once prepared with a bank of dispersed pulses, ESAM will evaluate the dispersion measure-time transform. 

The evaluation procedure proceeds in essentially a bottom-up fashion. Each EndNode is presented with a time series of size $N_t$ from its corresponding channel (data can be supplied directly to the end nodes, or by recursively splitting a block from the root node downwards). The EndNode performs a 1-D convolution of the input time series with each of its convolution kernels saved in its lookup table (list of products). It returns a block of convolution results with shape ($N_{prod}, N_t$) to its parent IterNode. 

Each IterNode now has convolution results from its upper and lower child nodes. The IterNode loops through its lookup table (which is a list of IterProducts) and computes a new output for each element in its lookup table. This output is the sum of the output from the lower output, and the upper output, with respective subband offset applied to the lower output. The outputs to be added, as well as the offsets to be used, are chosen based on the ProductIDs and subband\_offsets present in the IterProducts of that IterNode. 

 The IterNode evaluation pseudocode is shown in Listing~\ref{listing:eval_iternode}, and the EndNode evaluation pseudocode is in Listing~\ref{listing:eval_endnode}.

If the tree has been prepared with a bank of dispersed pulses, the root node computes and returns an ($N_{prod}, N_t$) shape array, which represents the dispersion transform of the input data. Each trace supplied in the preparation step will be reproduced with 100\% accuracy in the order supplied. No redundant operations are performed.

\begin{listing}[!ht]
\begin{minted}{python}
class IterNode:
    def prepare(self, trace):
        return prepare_iternode(self, trace)
    def eval(self, data):
        return eval_iternode(self, data)

class EndNode:
    def prepare(self, trace):
        return prepare_endnode(self, trace)

    def eval(self, data):
        return eval_endnode(self, trace)

\end{minted}
\caption{Types, showing how functions are dispatched}
\label{listing:types}
\end{listing}

\begin{listing}[!ht]
\begin{minted}{python}
def prepare_iternode(node, trace):
    n = len(trace.offsets)
    subband_offset = sum(trace.offsets[:n/2+1])
    upper_trace, lower_trace = split_trace(trace)
    upper_product_id = node.upper_child.prepare(upper_trace)
    lower_prouduct_id = node.lower_child.prepare(lower_trace)
    iter_product = (upper_product_id,
                    lower_produc_id, 
                    subband_offset)
    if product not in node.products:
        node.products.append(iter_product)
        
    product_id = node.products.get_index(iter_product)
    return product_id
\end{minted}
\caption{Preparation step of IterNode}
\label{listing:prep_iternode}
\end{listing}

\begin{listing}[!ht]
\begin{minted}{python}
def prepare_endnode(node, trace):
    if trace.kernel[0] not in node.kernels:
        node.kernels.append(trace.kernel[0])

    product_id = node.kernels.get_index(trace.kernel[0])
    return product_id
\end{minted}

\caption{Preparation step of an EndNode}
\label{listing:prep_endnode}
\end{listing}

\begin{listing}[!ht]
\begin{minted}{python}
def eval_iternode(node, data):
    nc = data.shape[0]
    nt = data.shape[1]
    # Get data from lower nodes recursively
    lower_data = node.upper_child.eval(data[:nc/2, :])
    upper_data = node.lower_child.eval(data[nc/2:, :])
    for (iprod, product) in enumerate(node.products):
        (upper_product_id, 
            lower_product_id, 
            subband_offset) = product

        if subband_offset > 0:
            output[iprod, 0:nt-subband_offset] =
            lower_data[upper_product_id, :nt-subband_offset]+
            upper_data[lower_product_id, subband_offset:nt]
        else:
            output[iprod, 0-subband_offset:nt] =
            lower_data[lower_product_id, :nt+subband_offset]+
            upper_data[upper_product_id, 0-subband_offset:nt]

    return output
\end{minted}
\caption{Evaluation step of an IterNode}
\label{listing:eval_iternode}
\end{listing}

\begin{listing}[!ht]
\begin{minted}{python}
def eval_endnode(node, data):
    for iprod, kernel in node.products:
        output[iprod, :] = convolve(kernel, data)

    return output
\end{minted}
\caption{Evaluation step of an EndNode}
\label{listing:eval_endnode}
\end{listing}

%
%

\subsection{Tips for designing kernel banks}
\label{sec:tips}
ESAM offers enormous freedom in choosing arbitrary kernels to cover the search parameter space. Here we offer a few tips on choosing kernels that should result in high levels of re-use with controllable accuracy.

\begin{itemize}

\item Quantise kernel weights: quantising the kernel weights to a small set of values increases the efficiency of re-use of EndNode outputs. The simplest (and most efficient case) is 1-bit (or binary) weights. By choosing as few bits as possible for the quantisation, 1-D Kernel weights are much more likely to be identical between different masks in the bank. Having identical 1-D kernels shared between masks in a bank means the same convolution output can be re-used for multiple top-level products. Binary masks also have the advantage that 1-D convolution can be efficiently implemented as a moving average filter, requiring only 2 additions and no multiplications per output point. By contrast, arbitrary weight convolution with a 1D kernel length $N$ requires $N$ multiply-adds per output point.

\item Evaluate the ESAM bank before adding new masks. An ESAM application will typically enumerate a set of desired masks and load them into a tree. The number of operations in the tree can be reduced if the user evaluates the ESAM tree on a proposed new mask before loading it. There will be cases where the existing result has sufficient S/N (according to some user-specified threshold), such that the proposed mask does not need to be added to the tree, and the additional operations needed to compute this mask exactly can be avoided. This approach is often used in designing search pipelines with brute force algorithms such as {\sc HEIMDALL}. This allows the users of ESAM to optimise the compute cost of the algorithm while staying above a desired accuracy threshold.

\end{itemize}

\section{Performance of ESAM}
\label{sec:ESAM_perf}

We evaluate the performance of the ESAM algorithm by testing the recovered signal-to-noise ratios for a range of simulated pulses, as well as counting the number of operations required. We describe our simulation technique, as well as the derived results, in the following subsections.


\subsection{Preparing the ESAM tree}

\subsubsection{Pulse simulation technique}
\label{subsec:pulse simulation}
To prepare an ESAM tree, we first create a set of dispersed pulses. We simulate dispersed radio pulses with narrow (delta-function) width, and dispersed following the dispersion equation \ref{Eq:dispersion eqn}. The signal fluence is kept uniform across all frequency channels. The amplitude of the pulse in each time sample is analytically computed based on the amount of time the pulse spends in a given sample (assuming that the pulse exits the lowest frequency channel in the middle of a sample). This allows for accurate reproduction of the effect of intra-channel dispersion, which is an important factor when evaluating the performance of a dedispersion algorithm. As a result of our choice of constant fluence, the matched filter S/N is not constant as a function of DM (the S/N decreases as the pulse amplitude gets smeared across multiple time samples at higher DMs).

The bank of pulses is simulated using the parameters shown in Table~\ref{tab:simparams}. These parameters are similar to those used in the ASKAP FRB localisation mode.\citep{Bannister2019}.

\begin{listing}[!ht]
\begin{minted}{python}
import numpy as np
def mfsn(signal, template):
    ''' 
    Returns matched filter signal to noise given a template
    '''
    return sum(signal*template)/sqrt(sum(template**2))

def quantise(mask):
    ''' 
    1-bit quantises input and returns float
    '''
    return (mask > 0).astype(float)
    
def make_esam_tree(rootnode, sn_threshold, dm_step):
    ''' 
    Populates an ESAM tree with FRBs given a S/N threshold
    and dm_step
    '''
    product_ids = []
    scales = np.array([])
    dm_range = np.linspace(dm_start, dm_end, dm_step)
    for ii, dm in enumerate(dm_range):
        frb = make_frb(dm)
        #quantise masks to 1-bit values to improve re-use
        proposed_mask = quantise(frb) 
        if ii == 0:
            #Always feed in the first mask
            pid = rootnode.prepare(proposed_mask)
            #save noise scaling as the square-root of
            #quadrature sum of the weights
            scales.append(sqrt(sum(proposed_mask**2)))
            product_ids.append(pid)
            continue

        #The matched filter S/N is obtained when the 
        #signal is convolved with a template that is
        #identical to the signal itself 
        theoretical_sn = mfsn(frb, frb)

        #We define the best possible S/N that can be
        #recovered as the S/N produced by the 
        #convolution of the signal and a template that
        #is the quantised version of the signal
        possible_sn = mfsn(frb, proposed_mask)

        #we calculate the achieved S/N by running
        #the ESAM tree we scale the output by the
        #inverse square of the weights
        achieved_sn = rootnode.eval(frb)/scales
        best_achieved_sn = output_sn.max()

        if achieved_sn/possible_sn < sn_threshold:
            #If the existing traces in the tree cannot 
            #recover the desired S/N then, save the 
            #proposed mask
            proposed_trace = mask_to_trace(proposed_mask)
            pid = rootnode.prepare(proposed_trace)
            scales.append(sqrt(sum(proposed_mask**2)))
            product_ids.append(pid)
    return product_ids

\end{minted}
\caption{Code used for preparing an ESAM tree with S/N threshold and DM step}
\label{listing:loader}
\end{listing}

\begin{table}
    \centering
    \begin{tabular}{rl}
       Bottom Frequency  & 800~MHz\\
       Channel width   & 1~MHz\\
       Number of channels  & 256 \\
       Integration time  & 1~ms\\
       Lower DM delay  & 0~ms \\
       Highest DM delay  & 1000~ms\\
       DM delay step & 0.1~ms \\
       S/N loss thresholds & 0.9, 0.8 \\
    \end{tabular}
    \caption{ESAM simulation parameters}
    \label{tab:simparams}
\end{table}

\subsubsection{Preparing the ESAM tree}
To reduce the number of operations, we apply the techniques of section~\ref{sec:tips} as shown in Listing~\ref{listing:loader} while building the tree.
We use 1-bit quantised weights in the tree. We enumerate real-valued pulses from the lowest DM delay (0 ms) to the highest DM delay (1000 ms) with small increments, depending on the tree parameterization as described in Section \ref{subsec:algo evaluation}. We calculate the best possible S/N (which we also refer to as `Max snr') by quantising the proposed pulse to 1-bit masks, and calculating the matched filter S/N of the unquantised pulse against the quantised mask. Next, we evaluate the ESAM tree feeding in the real-valued pulse as data. After evaluating against all masks loaded in already, the ESAM tree yields the achieved S/N. If the ratio of the achieved S/N to the best possible S/N is less than the supplied threshold, the trace is added to the tree. If the ratio is above the threshold, it indicates that the masks already loaded in the tree can recover the new pulse with S/N greater than the desired accuracy threshold from the user, hence the mask corresponding to the given pulse can be skipped to minimise the compute cost. If not, then the mask is converted to a trace and added to the bank of lookup tables stored in the tree.

\subsection{Measuring ESAM S/N recovery}
\label{subsec:esam snr recovery}

\begin{figure}
    \centering
    \includegraphics[width=0.95\linewidth]{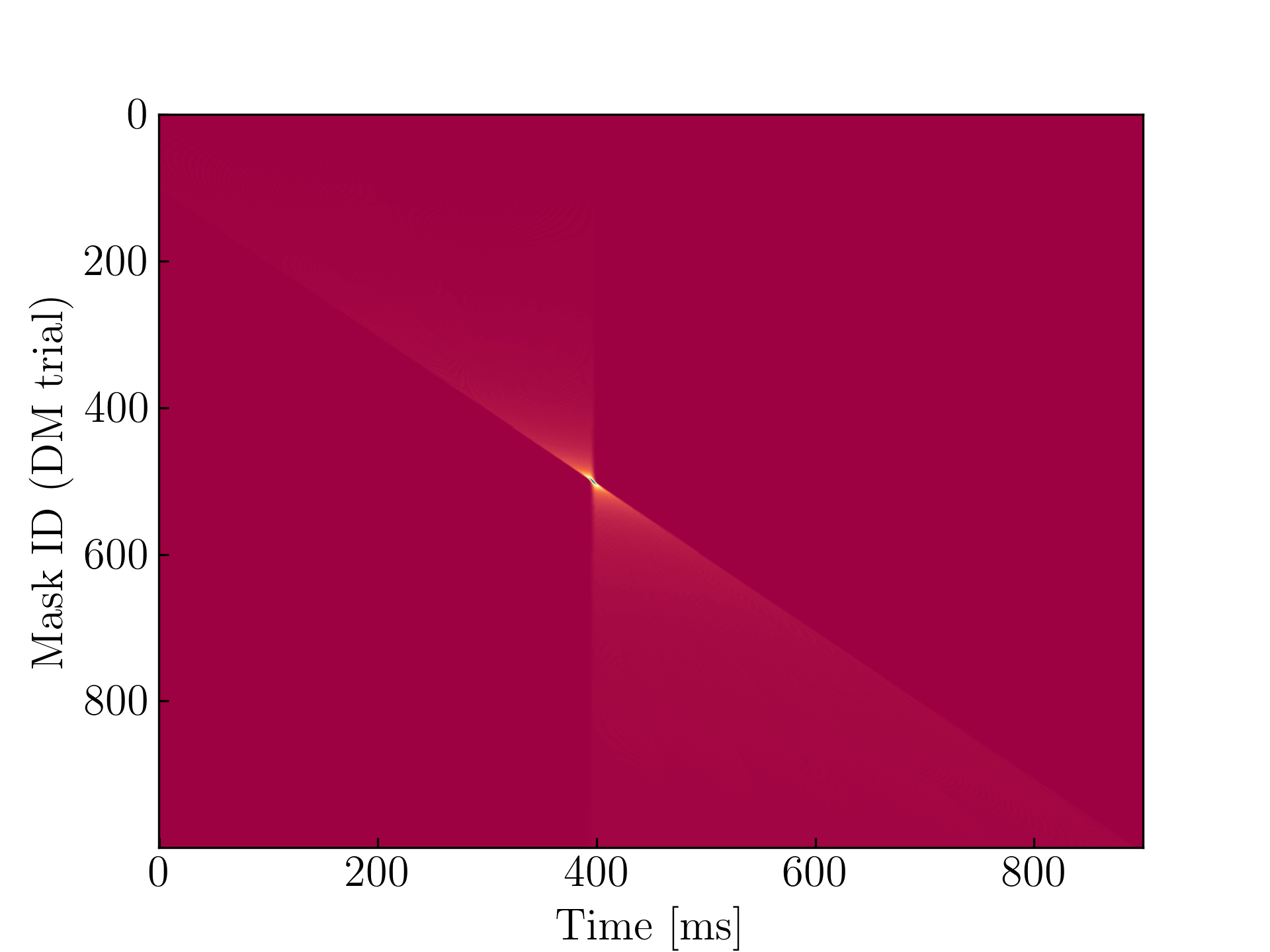}
    \caption{The output of ESAM tree evaluation for an example pulse dispersed at a DM delay of 500 samples. When loaded with dedipsersion masks, the ESAM produces the bow-tie pattern in its dispersion transform. }
    \label{fig:ESAM_bowtie}
\end{figure}

We create test dispersed pulses in the same way as described in Section \ref{subsec:pulse simulation} from the lowest DM delay (0 ms) to the highest DM delay (1000 ms) in increments of 0.1 ms, and evaluate the Dispersion Measure transform as described in Listing~\ref{listing:eval_endnode}. The output of the tree evaluation in ESAM gives us the sum of the signal along multiple mask and time trials. To obtain the recovered S/N, we divide the output with appropriate scales for each input mask, similar to what is described in Listing~\ref{listing:loader}, and find the peak in the scaled output. The output of the evaluation of ESAM tree built with 1000 DM trial masks for an example pulse dispersed at a DM delay of 500 samples is shown in Figure \ref{fig:ESAM_bowtie}.

\subsection{Measuring ESAM operation count}
\label{subsec:esam operation count}
We estimate the number of operations in the ESAM tree as the sum of $N-1$ operations for every 1-D convolution kernel of length $N$ plus 1 operation for every sum across frequency in the ESAM tree. ESAM will run very differently on different computing architectures, so these operation counts are a guide for the purposes of comparing parameterisations of ESAM (see below). We have functions in our reference implementation which compute these counts. 

\subsection{Algorithm evaluation}
\label{subsec:algo evaluation}

We evaluated the performance of 3 parameterised versions of ESAM as described below and compare them with brute force dedipsersion and the FDMT. In the following sections $ESAM(t,s)$ is a tree populated by the algorithm in Listing~\ref{listing:loader} with an S/N accuracy threshold $t$, and simulated DM delay step size $s$ (in ms).

For each algorithm, we simulate dispersed narrow test pulses in the DM delay range 0$-$1000 ms with 0.1 ms increments, and compute the recovered S/N as well as the number of operations required.

{\bf Brute Force} --- For brute force, we calculate the S/N of a matched filter with 1-bit quantised masks made from the simulated pulses (described in Section \ref{subsec:pulse simulation}). The dedispersion masks are spaced at integer sample (1 ms) delay increments in the range 0-1000 ms. We compute the number of operations (sums) required as the number of non-zero values in each mask.

{\bf FDMT} --- we create an FDMT class (as described in \cite{Zackay2017_FDMT}) and evaluate it on the bank of simulated test pulses. To find the S/N of a simulated pulse, we find the maximum value in the DM-time transform returned by FDMT, and note that value's DM and time index. We then pass a frequency-time block where every value is set to 1, to the FDMT class, and find the output value at the same DM and time index. This gives us the sum of FDMT's implicit mask, i.e. the number of samples added by FDMT. We divide the maximum value in the DM-time transform with the square root of the sum of the mask to get the signal-to-noise ratio recovered by FDMT\footnote{Ideally, we should take the square root of the sum of the squares of the weights used in FDMT's implicit mask. However, since all weights are 1, taking a simple sum also gives us the correct answer.}. 
To calculate the compute cost of FDMT, we extract the implicit masks the FDMT uses and load those kernels into an ESAM tree. We then use ESAM to count the number of operations the ESAM tree loaded with FDMT masks will perform (see Section \ref{subsec:esam operation count}). We have checked that this ESAM tree performs an identical transform to the FDMT.

{\bf ESAM$(1,1)$} --- We build an ESAM tree with 1-bit quantised masks of simulated pulses distributed with a DM step of 1 ms and a threshold of 1 (i.e. all masks get loaded in the tree). This results in a tree with the same number, and distribution of DM trials as the brute force algorithm and the FDMT (i.e. integer sample spaced DM trials), but has traces generated from simulated pulses, as distinct from the implicit traces used by the FDMT.

{\bf ESAM$(0.9,0.1)$} --- We build an ESAM tree with 1-bit quantised masks of simulated pulses with a DM step of 0.1 ms and a threshold of 0.9. This illustrates the tunability of ESAM to balance computational cost and accuracy.

{\bf ESAM$(0.8,0.1)$} --- ESAM tree built with 1-bit quantised masks of simulated pulses with a DM step of 0.1 ms and a threshold of 0.8. This illustrates the improvement in computational cost one can get for a minimal hit in accuracy.

\section{Results}
\label{sec:results}

\begin{figure*}[htb!]
    \centering
    \includegraphics[width=0.75\linewidth]{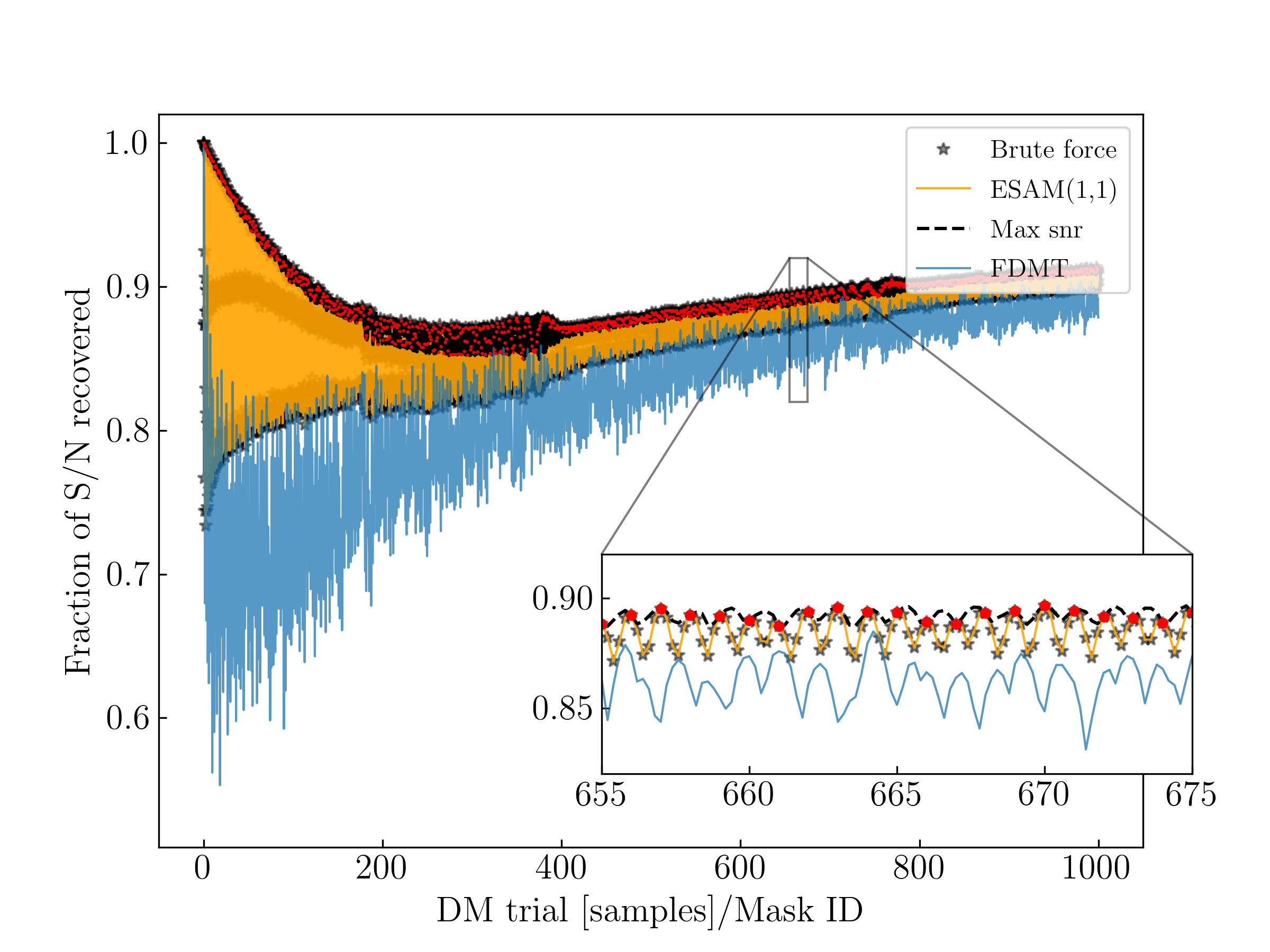}
    \caption{Recovered S/N as a fraction of the theoretical S/N evaluated for a range of algorithms and FRB DMs. The maximum recoverable S/N using quantised kernels is shown with a black dashed line. The performance of ESAM(1, 1) is shown in orange, and FDMT is shown in blue. The performance of the brute force algorithm is shown with black stars. ESAM(1, 1) exactly matches the performance of the brute force algorithm, while only requiring 10x fewer operations (see Figure~\ref{fig:compute_cost}). Red circles mark the performance of ESAM(1, 1) where the test DM trial matches the input mask, i.e. at integer DM trials. ESAM(1, 1) recovers the max S/N at those DMs. Inset: zoom-in on a small region of the three curves. }
    \label{fig:ESAM1_snr_recovery}
\end{figure*}

Figure \ref{fig:ESAM1_snr_recovery} shows the S/N recovery of ESAM$(1, 1)$ as a function of test DM trials (spaced at 0.1 ms delay increments). We also plot the S/N recovered by brute force and FDMT algorithms for comparison. The best possible recovered S/N is not uniformly 100\% across all DMs because of the choice of  1-bit quantised masks and constant fluence, rather constant S/N, of the simulated FRBs. The FDMT algorithm loses up to 40\% in S/N due at low DMs, due to the low mask accuracy.

The S/N recovered by the ESAM(1,1) exactly matches the performance achieved by the brute force algorithm. Since this tree is loaded with all masks spaced by 1 sample delay, it recovers the best possible S/N for all integer DM trials (red dots in the inset zoom plot). In the gaps between two adjacent trial DMs, the S/N recovery is lower, due to scalloping. Despite the fact that ESAM(1,1) has been built with the same number, and spacing, of masks as the FDMT, on average, the performance of ESAM(1,1) exceeds that of FDMT because of the higher accuracy of the masks loaded in the tree. 

\begin{figure}
    \centering
    \includegraphics[width=0.95\linewidth]{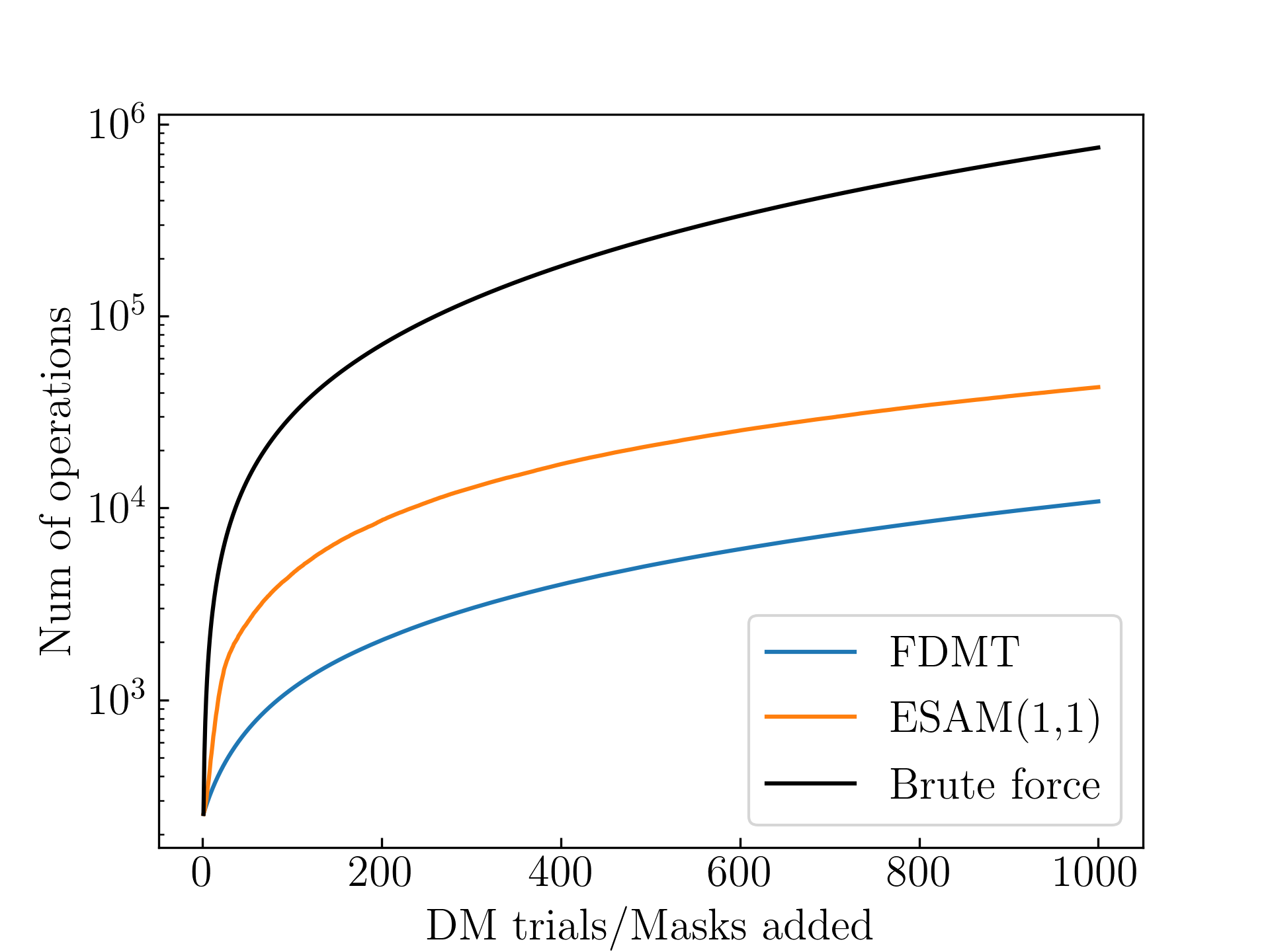}

    \caption{Number of operations needed to evaluate dedispersion for a range of algorithms and FRB DMs.}
    \label{fig:compute_cost}
\end{figure}

A plot of the number of operations needed by each algorithm is shown in Figure~\ref{fig:compute_cost}. While both brute force and ESAM(1, 1) recover the same S/N, brute force has a $\mathcal{O}(N_d,N_c)$ complexity and requires $>10\times$ more operations to perform the same operation. Both ESAM(1,1) and FDMT reuse partial sums and lower their computational complexity to $\mathcal{O}(N_d, log_2 N_c)$. However, ESAM(1,1) only reuses fully redundant partial sums, and remembers all unique operations that need to be performed to accurately reproduce the bank of masks specified by the user. This results in ESAM needing to perform more operations than the FDMT, which has a higher reuse of partial sums but at the cost of mask accuracy. 

ESAM is inherently a generic algorithm. If an ESAM tree is built with finely spaced DM trial masks, its S/N performance would become even better, just like that of the brute force algorithm (with the same finer spacing of DM trials), at increased compute cost but still maintaining the lower computation complexity resulting from the efficient reuse of partial sums. 
Conversely, if an ESAM tree is built with masks that match the masks implicitly used by the FDMT algorithm, then that tree would have identical computation cost and performance as the FDMT (the orange line would overlap with the blue line in Figures \ref{fig:ESAM1_snr_recovery} and \ref{fig:compute_cost}).

All algorithms show significant loss of S/N between adjacent DM trials due to scalloping. This problem can be tackled by increasing the resolution of the DM search trials. Increased DM resolution leads to a proportionate increase in the compute cost.
The generality in the ESAM algorithm allows us to easily build trees optimised for better S/N accuracy without necessarily increasing computational requirements. We demonstrate this optimisation in the next section.

\subsection{Optimised ESAM trees}
\label{subsec:optimised_esam_results}

\begin{figure}
    \centering
    \includegraphics[width=0.95\linewidth]{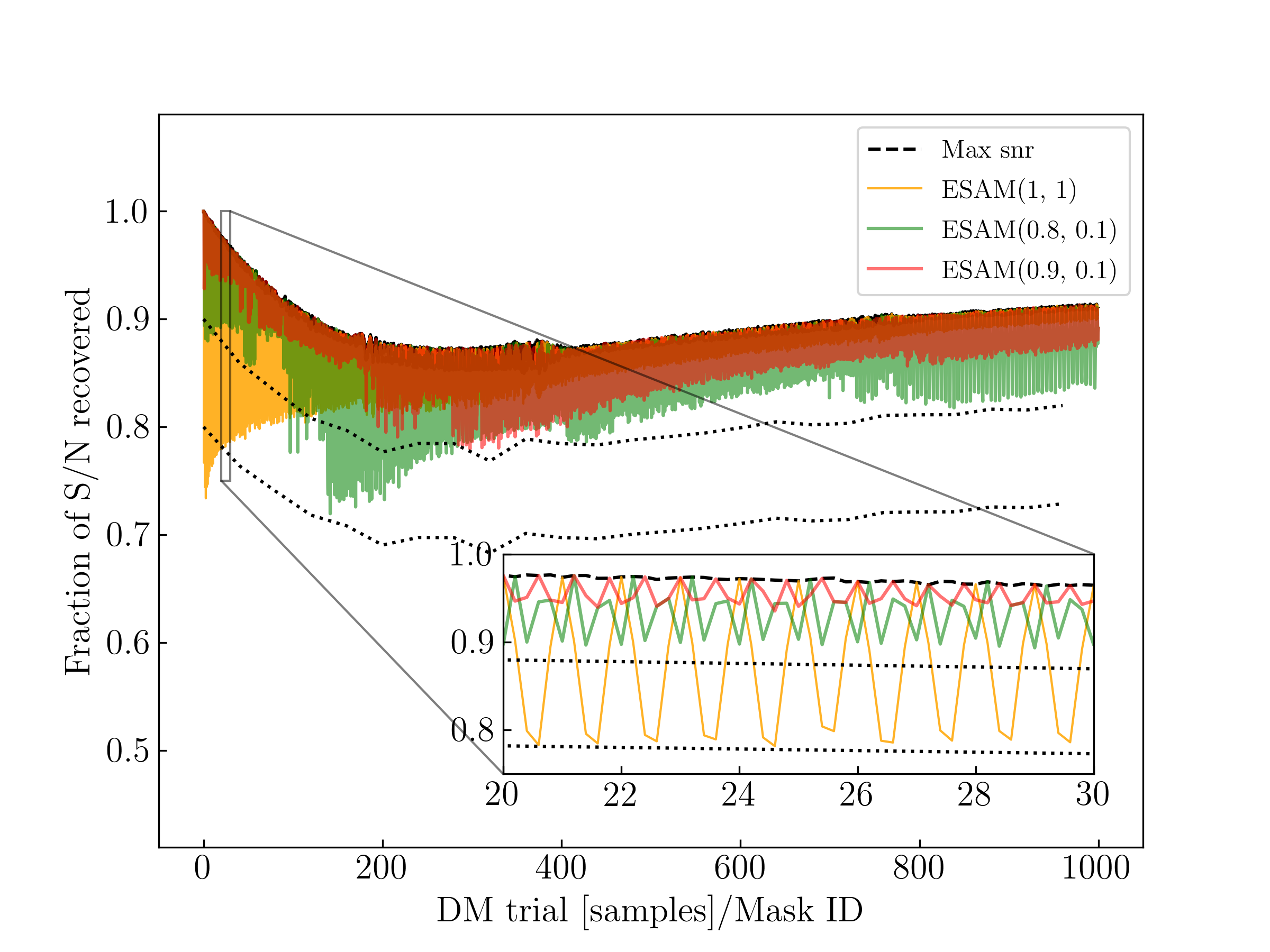}
    \caption{S/N recovery performance of the optimised ESAM trees $-$ ESAM(0.9,0.1) in red, and ESAM(0.8,0.1) in green, as a function of DM. The best possible S/N is shown in the black dashed line, and the performance of ESAM(1,1) is shown in orange for comparison. Black dotted lines show the 90\% and the 80\% S/N recovery thresholds used to build the two optimised trees. Both trees exceed the performance of the ESAM(1,1) tree in the low DM trial region --- where the latter suffered. Their performance drops slightly at high DM trials but stays within their specified thresholds. This results in a reduction in the number of operations needed, as shown in Figure~\ref{fig:optimised_compute_cost}.
}
    \label{fig:ESAM_snr_comparison}
\end{figure}

\begin{figure}
    \centering
    \includegraphics[width=0.95\linewidth]{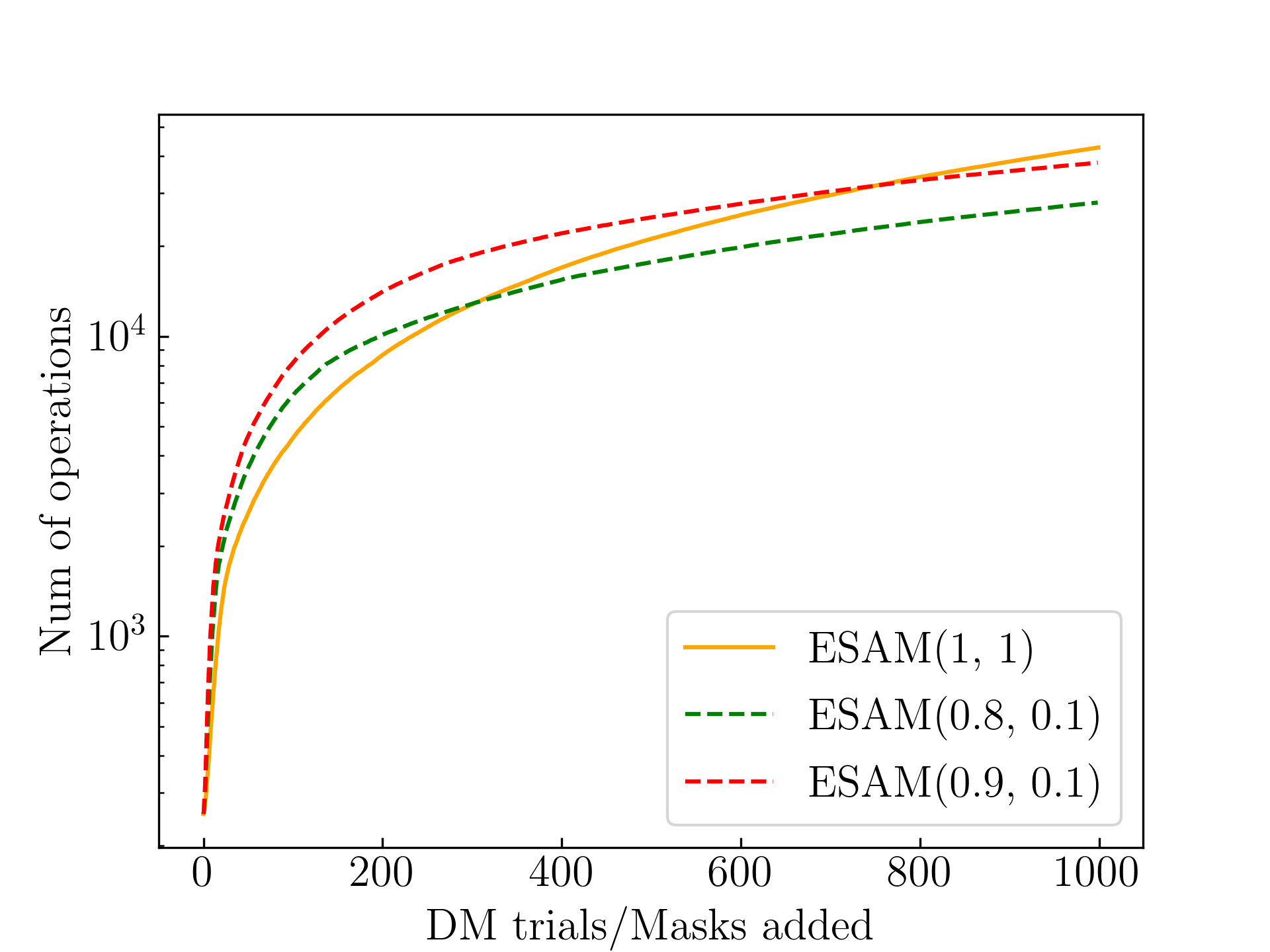}

    \caption{Number of operations needed to evaluate the dedispesion transform for ESAM trees with differing parameters. }
    \label{fig:optimised_compute_cost}
\end{figure}

We have built an ESAM tree --- ESAM(0.9,0.1) which guarantees $>90\%$ S/N recovery at all DMs. As demonstrated in Listing \ref{listing:loader}, and builds the tree with masks with a fine DM spacing (0.1 ms) between them. The S/N performance of this tree is shown in Figure$~$\ref{fig:ESAM_snr_comparison}. The tree is clearly able to recover $>90\%$ of the maximum possible S/N. Importantly, the performance of ESAM(0.9,0.1) is better than that of ESAM(1,1) at low DMs --- where ESAM(1,1) was missing more than $10\%$ of the S/N. The compute cost of ESAM(0.9,0.1) is shown in Figure$~$\ref{fig:optimised_compute_cost}. Overall, the number of operations needed remains in the same order of magnitude as ESAM(1,1). However, the tree requires more operations at low DMs and has reduced requirements for higher DMs. This explains how the tree is able to recover better S/N at low DMs, but loses more S/N at higher DMs than ESAM(1,1). 

For illustrative purposes, we have also shown results for another tree --- ESAM(0.8, 0.1) with a lower S/N accuracy threshold. As evident from Figure$~$\ref{fig:ESAM_snr_comparison} and Figure$~$\ref{fig:optimised_compute_cost}, this tree achieves the desired S/N threshold with further relaxed computational requirements. We encourage the readers to build their own custom tree optimisation schemes that suit their needs.


This demonstrates that the designer can, with minimal effort, trade between computational cost and recovery S/N. This allows designers to eliminate the problem in FDMT of inefficient distribution of DM trials for a sensitive search \citep{Rajwade2024_frb_search_algo_review}.

\section{Discussion and conclusions}
\label{sec:discussion}

We have introduced a new algorithm ESAM, based on the divide-and-conquer strategy, capable of efficiently computing 1-D convolutions of arbitrary 2-D masks on 2-D data arrays. We demonstrate how this method can be used to construct a dedispersion engine that achieves high accuracy while maintaining lower computational complexity compared to brute-force implementations. We find that ESAM has a similar S/N performance to a brute force algorithm while requiring an order of magnitude fewer operations. We note that our analysis here relies on the total operation count for each algorithm. We do not compare the actual runtime of the algorithms, which can depend on the individual implementation. For example, the GPU-accelerated implementation of the brute force algorithm in HEIMDALL will likely have a lower runtime than our vanilla Python implementation of ESAM, despite having a much higher operation count. Developing parallelised/GPU-accelerated implementations of ESAM is left for future work.

This algorithm has potential applications beyond the dedispersion case. For any application where a 1-D convolution of a large number $(\gg 1)$ of 2-D masks with data is required, this algorithm can be used to compute the convolution efficiently and accurately. For instance, searching for technological signatures in dynamic spectra from radio telescopes involves convolving data with multiple trials of Doppler delays \citep{Enriquez2017_BL_search_692stars}. An ESAM tree can be built using simulated masks of all the desired Doppler delays within the search window, in an identical fashion to the way we built our tree for multiple DM trials in Section~\ref{sec:ESAM_perf}. Since the tree is generated directly from the simulated masks, there is no need to explicitly code the analytic equations that parameterise the necessary transform for each trial. 
This flexibility also allows designers to incorporate nuanced features into their searched templates, such as the frequency-dependent pulse broadening due to the effect of interstellar scattering, or temporal drifting of signal to lower frequencies (the `sad trombone' effect) seen in some repeating FRBs --- features that are easier to simulate, but challenging to code in inversion transforms.

We acknowledge that, unlike other dedispersion algorithms that function with minimal setup `out of the box', ESAM requires users to invest time in preparing the ideal set of masks for their data and building the corresponding ESAM tree. However, this additional preparation step makes ESAM a versatile and adaptable algorithm, offering the flexibility to tailor it to specific use cases, leading to improved S/N accuracy and lower compute costs. Once the tree is built, it can be saved to disk as a set of lookup tables or as a serialized Python object (if implemented in Python), making it easy to share across different users and platforms. For demonstration purposes, we have provided an example ESAM(1, 1) tree as a pickled Python object in the GitHub repository (https://www.github.com/vivgastro/ESAM).




\begin{acknowledgement}

The authors are grateful to Dr. Pravir Kumar, Dr. Barack Zackay and Dr. Daniel C. Price for useful discussions during the preparation of this manuscript. This project was supported by resources and expertise provided by CSIRO IMT Scientific Computing. This research made use of {\sc numpy} \citep{numpy}, {\sc pandas} \citep{pandas-official}, {\sc matplotlib} \citep{matplotlib} and {\sc jupyter} \citep{jupyter} packages.

\end{acknowledgement}

\bibliography{big_bibliography, python_packages}

\appendix

\end{document}